# Exploring Channel Probing to Determine Coherent Optical Transponder Configurations in a Long-Haul Network

Kaida Kaeval[1], Danish Rafique[1], Kamil Bławat[1], Klaus Grobe[1], Helmut Grießer[1], Jörg-Peter Elbers[1],
Piotr Rydlichowski[2], Artur Binczewski[2], Marko Tikas[3]

[1] *ADVA Optical Networking SE, Martinsried, Germany*
[2] *Poznan Supercomputing and Networking Center, Poznan, Poland*
[3] *Tele2 Estonia, Tallinn, Estonia*
kkaeval@advaoptical.com

**Abstract:** We use channel probing to determine the best transponder configurations for spectral services in a long-haul production network. An estimation accuracy better than ±0,7dB in GSNR margin is obtained for lightpaths up to 5738km. © 2020 The Author(s).
**OCIS codes:** 060.2330 Fiber Optics Communications; 060.2360 Fiber Optics Links and Subsystems; 060.1660 Coherent communications

## 1. Introduction

Facilitated by flexgrid-capable reconfigurable optical add/drop multiplexers (ROADMs) and digital signal processor (DSP) enabled coherent transponders, optical Spectrum as a Service (OSaaS) ideas are gaining popularity in dense wavelength division multiplexing (DWDM) networks [1]. OSaaS allows operators to share unused fiber capacity with customers and other operators, thereby increasing network utilization and reducing infrastructure costs.

To characterize a spectral slot on an optical lightpath, a measure for the attainable transmission performance is required. While the optical signal-to-noise ratio (OSNR) is a common performance criterion, it does not consider the impact of fiber nonlinearities. To account for these effects, the use of a generalized OSNR (GOSNR) was proposed [2]. Unfortunately, the GOSNR cannot be directly measured; it needs to be calculated from optical system parameters or inferred from indirect measurements. For subsea open cable systems, it has recently been suggested to use a test transponder to probe a subsea link and to derive the GOSNR distribution over wavelength from this measurement [3].

In this paper, we follow a similar approach for terrestrial long haul network and determine its applicability in a practical network context. We probe the optical channel with a transponder in a fixed reference configuration to estimate the GOSNR of a spectrum service in a long-haul network. By normalization of GOSNR to the symbol rate, a modulation format independent generalized signal-to-noise ratio (GSNR) is obtained. From the GSNR we then predict the best transponder configuration for different spectrum services delivering maximum capacity at highest spectral efficiency. We verify the accuracy of the method in Tele2 Estonia's terrestrial long-haul network between Tallinn and Frankfurt for lightpath lengths up to 5738km. A commercial coherent transponder offering highly flexible 100G-600G dual-polarization quadrature amplitude modulation (DP-QAM) modulation is used both for channel probing as well as for data transmission. Compared to previous work on automatic transponder configuration [4], we investigate the practicability of channel probing and configuration optimization on a large set of long-haul distances and modulation formats. We show that false positives can occur in predicting the best transponder configuration. Margins and/or an additional optimization step can be applied to solve this issue.

## 2. Method description

Following a back-to-back characterization of the probing light transponder (PLT), the channel probing method consists of two steps: Estimation of the channel performance in a spectrum slot by means of the GSNR derived from a reference measurement and selection of best modulation format/symbol rate combination for that lightpath. A back-to-back measurement of the Q-factor versus the OSNR for at least one modulation format/symbol rate combination is required to characterize the PLT. The logarithmic Q-over-OSNR characteristic is then fitted by a 2nd order polynomial and the resulting function is then inverted to estimate the OSNR from the Q-factor. For our tests, the Q-over-OSNR characteristics were taken with one PLT module, and the actual network probing was conducted with a second module, with no further cross-characterization between the modules.

In order to evaluate the actual channel performance of a lightpath at hand, the selected probing-light is inserted into the network in the respective spectral slot and the Q value is read out from the receiver to estimate the effective GOSNR. The GOSNR considers all optical distortions that impact the optical signal, including ASE noise, nonlinear distortions, etc. as well as any transceiver impairments. This value is then normalized to the symbol rate of the probe signal to obtain the GSNR. As second step, this GSNR is then compared to the typical GSNR values of different modulation format/symbol rate combinations as per system specification. Note that specified typical vs. worst-case GSNR can vary, which must be considered when looking at estimations leading to close-to-zero margin. Margins can be added to the GSNR estimates to increase the robustness of the prediction to any fluctuations and ageing effects in the network [5].

## 3. Network and transceiver configurations for the verification tests

A coherent flex-grid network with a maximum physical path length of 2869 km and more than 6 Tbit/s of live traffic in service was used to investigate the feasibility of the method. Several optical lightpaths with lengths up to 5738 km, enabled by optical loopbacks in intermediate ROADM nodes, were investigated. We used four 100-GHz spectrum slots with the central frequencies 193.90 THz for the shortest lightpath 194.00 THz for a 1792-km path, 194.10 THz for a 2943-km path, and 194.20 THz for the three longest paths, respectively (see Figure 1). A passive 4×1 splitter/combiner was used in the add/drop path to insert and extract the signals.

After inserting the selected PLT into the respective links and estimating the GSNR, verification signals with different modulation format/symbol rate combinations were tested, keeping the power spectral density (PSD) constant for all signals under test. Then, margins were calculated between the PLT-estimated GSNR and typical specified GSNR per used modulation format/symbol rate as

$$\text{Margin} = \text{estimated GSNR} - \text{typical, specified GSNR} \qquad (1)$$

These margins were compared with the actual performance of the verification signals. If the margin estimation was positive, the verification signal was expected to work, leaving non-working channels as false positive results.

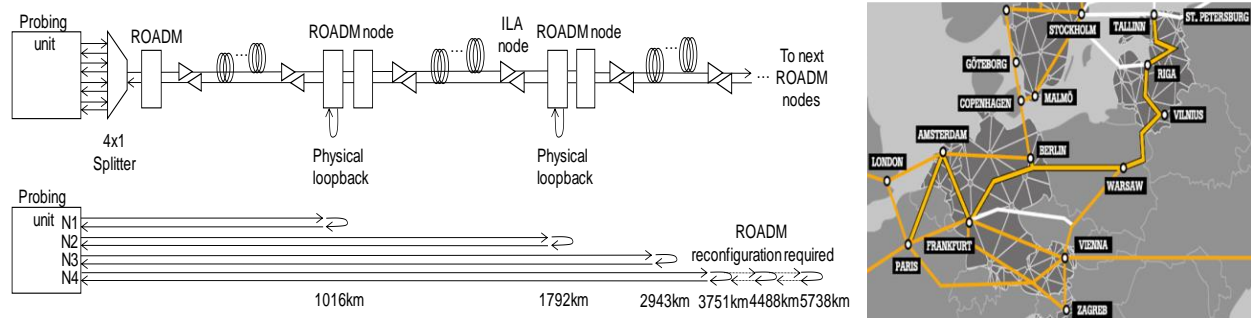

Figure 1. Test schema and network layout.

Although all conditions were held as stable as possible during the tests, certain variations in OSNR and polarization cannot be avoided in a live network with lightpaths of more than 5500 km in length. As such, we do not expect our method to deliver a perfect solution but rather a practical estimation.

## 4. Results and interpretation

We demonstrate the most important findings of our work in Figure 2. The main diagram shows the estimated GSNR margin on vertical axes over the different tested link lengths. All link performance estimations for the main diagram were retrieved using 200Gbit/s 69GBd DP-QPSK probing light that provided the best estimation accuracy over the other tested probing settings. Verification signals were selected with 0.5bit/symbol steps and 100G/200G/300G/400G data rates were tested. Different line styles refer to different data modulation formats/symbol rates and different colors to different line rates. The best transponder configuration for a given lightpath is the signal with the lowest GSNR margin, which is still working. As expected, the estimated GSNR margin decreases with increasing link length.

An estimation accuracy of better than ±0,7dB in GSNR margin is achieved for lightpaths up to 5738km. For low estimated GSNR margins some false positive results were estimated. These recorded estimation errors can be attributed to parameter changes in the photonic layer over the test period, slightly deviating performance of the PLT unit used in the tests compared to the characterized one, and the typical GSNR values used in Eqn. (1). The estimation errors were spread across all measurements, and no obvious correlation with modulation format, symbol rate or signal PSD was observed. For practical use, an additional operating margin would eliminate most of these errors. An additional fine-tuning step could also be applied to optimize the transponder configuration in the region where the estimated GSNR is low.

We also compared the link GSNR estimation accuracy when varying the symbol rates and constellation sizes of the probing light. In particular, the GSNR was estimated for: 34-GBd 100-Gbit/s DP-QPSK (PL1), 69-GBd 200-Gbit/s DP-QPSK (PL2), 69-GBd 300-Gbit/s DP-8QAM (PL4), 69-GBd 400-Gbit/s DP-16QAM (PL4) probing signals. The results of these comparisons are presented as inserts for Figure 2. In these inserts, the shortest and longest links were chosen where all four probing light settings gave meaningful results, so A and B reflect a transmission distance of 1016km and 1792-km, respectively. We can see that the 34GBd 100Gbit/s DP-QPSK signal has the tendency to overestimate the lightpath GSNR, which leads to false positive results, while higher modulation formats are prone to estimation errors, most likely caused by the fact that higher constellations are more prone to noise. Overall, the inserts clearly show that the overall performance of the highest symbol rate with lowest constellation size (200Gbit/s 69GBd DP-QPSK) gave the best result.

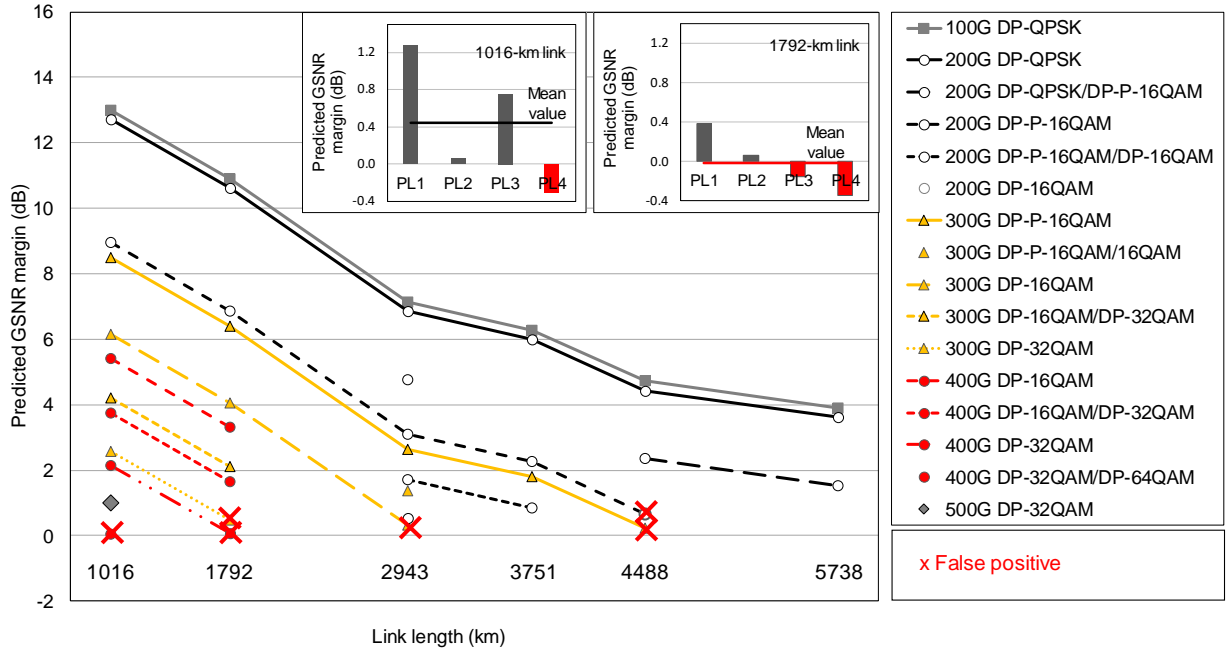

Figure 2. Predicted GSNR margin for different verification signal combinations with 200-Gbit/s PLT. The inserts A. and B. show the estimation deviation of the four different PLT settings.

## 5. Conclusions

Channel probing – including using data from already commissioned transponders – can be a cost-efficient approach in particular for quick link performance estimations. It can help eliminating uncertainty in simulations or calculations, which practically cannot consider all effects in reconfigurable and time-dependent networks.

In this paper, we investigated the performance of channel probing on a long-haul commercial network and showed consistent results for different symbol rates, modulation formats and link length combinations. Using a 200Gbit/s probing-light signal, an estimation accuracy of better than ±0,7dB in GSNR margin is achieved for lightpaths up to 5738km. For low estimated GSNR margins some false positive results were estimated. The errors fell well within both, the maximum network variations and the difference between typical vs. worst-case specifications.